\documentstyle[prl,multicol,aps,psfig,epsf]{revtex}
\begin{document}
\draft
\title{Tricritical behaviour of Ising spin glasses with charge fluctuations}
\author{B. Rosenow$^1$ and R. Oppermann$^{1,2}$}
\address{$^1$ Institut f\"ur Theoretische Physik, Univ. W\"urzburg, D--97074 
W\"urzburg, FRG} 
\address{$^2$ Service de Physique Theorique, CE Saclay,
F-91191 Gif-sur-Yvette, France}
\date{January 30, 1996}
\maketitle
\begin{abstract}
We show that tricritical points displaying unusal behaviour exist in 
phase diagrams of fermionic Ising spin glasses as the chemical potential
or the filling assume characteristic values. Exact results for
infinite range interaction and a one loop renormalization group analysis 
of thermal tricritical fluctuations for finite range models are
presented. Surprising similarities with zero temperature transitions and a new $T=0$ tricritical point of 
metallic quantum spin glasses are derived. 
\end{abstract}
\begin{multicols}{2}
\narrowtext
\tighten
Spin glass phase transitions can have important effects on a variety of 
characteristics of fermionic systems at low temperatures including 
nonmagnetic properties, and vice versa. This has been emphasized in 
a series of recent theoretical articles on quantum spin glass transitions 
\cite{Sachdev1,Sachdev2,Georges,RSY,SRO}. Considerable  
interest in these problems was raised by experimental results for 
heavy fermion systems \cite{Maple,Loehn}.
Spin glass phases known to exist between antiferromagnetic-- and 
superconducting phases of LaSrCuO phase diagrams \cite{LaSrCuO} may 
also be 
considered under the aspect of interplay of magnetic and electronic properties.
\\
\indent
In this Letter we report results on existence and properties of 
tricritical fermionic spin glass transitions which happen to emerge as the
fermion concentration, relevant for the effective spin dilution, is moved 
through a characteristic value. 
We analyzed in detail the tricritical behaviour of the
Ising spin glass (denoted by $ISG_f$) on fermionic
space with 4 states per site instead of the usual two
of SK models for example. Results are given i) for {\it
infinite-range-},
and (by use of the renormalization group) ii) for {\it
finite-range} spin interaction, and iii) for a
{\it metallic model} with additional electron hopping
hamiltonian.
The two nonmagnetic states per site provided by the 
fermionic space allow the system to adjust its effective random spin 
dilution according to quantum statistics. On decreasing the effective spin
density and hence $T_c$ the system is driven through a tricritical
point into a regime of discontinuous spin glass transitions. The 
tricritical point ({\it TCP}) turns out to be particularly interesting, 
since 
quantities such as density of states, fermion concentration (for given 
$\mu$ and vice versa), local susceptibility, and spin correlation function 
behave nonanalytically at the tricritical spin glass transition, and
thereby change substantially critical properties of those quantities 
which typically define and signal spin glass transitions.
Experimental observation is also favoured by several aspects
such as
increased upper critical dimension 
$d^{(u)}_c=8$ 
(instead of a $d^{(u)}_c$ decreasing from 4 to 3 in conventional 
$\phi^6$--theories \cite{LawSa}), which allows critical fluctuations in 
$3D$ to have 
much stronger effects than the usual logarithmic corrections
and the occurence of phase separation in the 1st order
regime, a phenomenon shared by the phase diagram of the
BEG-model for $He^3-He^4$ mixtures \cite{BEG}. We also
compare with other classical spin--1 models
\cite{MS}.\\
\indent
Understanding the key features of the $ISG_f$
as the simplest model which takes spin glass order and
charge correlations into account provides a
useful guide to phase diagrams of spin glasses allowing
for thermally activated hopping or metallic
conductivity. It appears to be generic for the
behaviour of an even larger class of models, in a way
comparable with the BEG--model \cite{BEG}.
The $ISG_f$ is defined by the hamiltonian 
${\it H}=-\frac{1}{2}\sum J_{ij}\sigma_i\sigma_j-\mu\sum
n_i$ 
with spins $\sigma =n_{\uparrow}-n_{\downarrow}$, the
fermion number operator $n=n_{\uparrow}+n_{\downarrow}$ 
, and gaussian--distributed 
$J_{ij}$ with variance $J$. All spin-- or charge--correlations of this 
model are static.
As an example the local susceptibility $\chi_{ii}(t)$ has the Fourier 
transform $\chi(\omega)=\beta(\tilde{q}-\int^{1}_{0} q(x))\delta_{\omega 
0}$ with $\tilde{q}=<\sigma_i(t)\sigma_i(t^{\prime})>$ and $q(x)$ 
denoting the Parisi solution of the $ISG_f$.
In contrast to the
standard $2$--states per site Ising spin glass this $4$--states per site model  
feels quantum statistics due to the relative occupation of magnetic 
and nonmagnetic states. Quantum dynamics however enters in the electron 
Greens function and 
correlations being defined with an odd number of equal--time fermion 
operators. The fermionic path integral representation not only produces 
the correct spin--static field theory for the spin glass transition, 
it also helped us to find a close
relationship between nonanalytical behaviour of the density of states or 
fermion concentration and the special features of the tricritical spin glass 
transition. In the same way we 
identify the surprising similarity between this classical tricritical 
theory for finite $T_c$ and the recently analyzed quantum theory of the 
$T=0$--quantum paramagnet to spin glass transition of a metallic model.
\\ 
\indent
\indent
Working in the grand canonical ensemble our detailed analysis of the $ISG_f$
shows that upon increasing 
the chemical potential from $\mu =0$ at half--filling 
the line of continuous spin--glass transitions 
given by $T_{c}=J\tilde{q}(T_{c})$ is only realized up to a tricritical
point located at $T_{c3}=J/3$. Beyond this point, ie for $T_c<\frac{J}{3}$
one enters the domain of discontinuous spin--glass transitions. 
Fig.\ \ref{fig1} displays this most interesting part of the phase 
diagram which 
surrounds the {\it TCP}. All relevant zero--field properties can be 
concluded from our result for the saddle--point free energy and 
expanded around the {\it TCP}, which reads (the reader
may also extract the tricritical behaviour from the
simpler approximate form using $q(x)=q_{EA}$ constant)
\begin{eqnarray}
f&-&f_{TCP}=\mu -\mu_{c3}+J\{(\frac{3}{2}r_g 
g-r_{\tau}\tau^2)\delta\tilde{q}
+\frac{3}{2}\tau[(\delta\tilde{q})^2\nonumber\\ 
&-&\int^{1}_{0}dxq^2(x)]
-\frac{3}{2}
[\int^{1}_{0}dx[xq^3(x)+3q(x)\int^{x}_{0}dyq^2(y)]\nonumber\\
&-&3\delta\tilde{q}\int^{1}_{0}dxq^2(x)
+\frac{1}{4}(\delta\tilde{q})^3]-\frac{y_4}{4}\int^{1}_{0}dxq^4(x)\}
\label{one}, 
\end{eqnarray}
where $\delta\tilde{q}\equiv\tilde{q}-\tilde{q}_{TCP}, 
gJ=\mu-\mu_{c3}+(\zeta^{-1}J-\mu_{c3})\tau$ as nonordering field, and 
$\tau\equiv\frac{T-T_{c3}}{T_{c3}}$. The constants are given by 
$r_g=\frac{2\zeta}{3}, 
r_{\tau}=2(1-\frac{3}{4}\zeta^{-2})$ 
with $\zeta\equiv tanh(\mu_{c3}/T_{c3})\simeq 0.9938$, and 
$\mu_{c3}=\frac{J}{3}arcosh(2exp(\frac{3}{2}))\simeq 0.9611J$ as the 
characteristic chemical potential locating the {\it TCP}.
The average filling factor corresponding to $\mu_{c3}$ is evaluated
as $<\nu_{c3}>\simeq 1.6625$. 
By symmetry w.r.t. $\mu=0$ one also 
finds a $TCP$ for less than half--filling at $<\nu_{c3}>\simeq 0.3375$. 
Thus the low-- and the high--filling domains host discontinuous spin 
glass transitions. We shall argue below that this is a rather general 
phenomenon.\\ 
\psfig{file=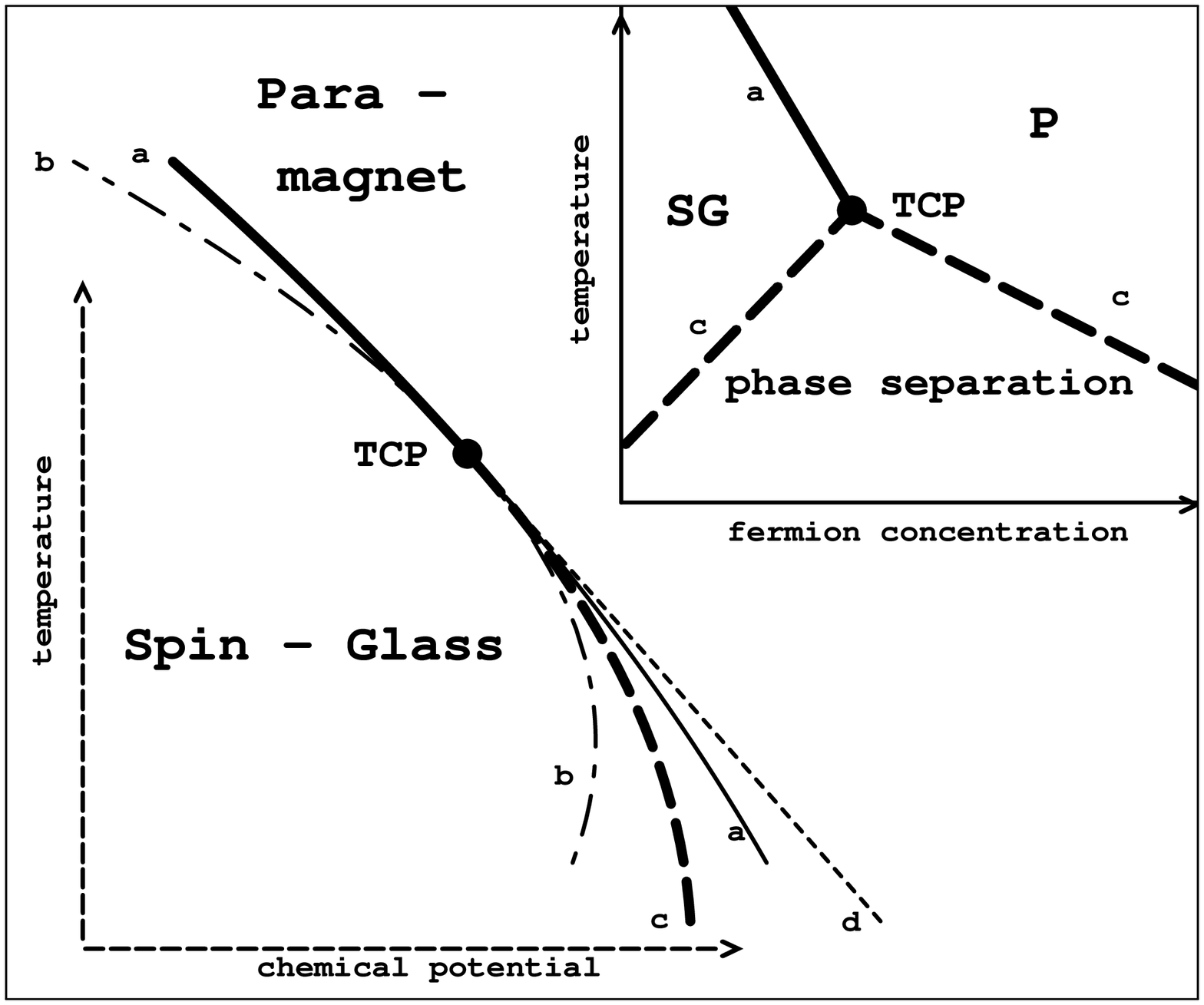,width=6cm,angle=90}
\begin{figure}
\caption
{Vicinity of tricritical point ($TCP$) for positive chemical
potential.
Continuous spin glass transitions occur on curve (a) above the $TCP$
(thick unbroken line).
Below the $TCP$ first order thermodynamic transitions take place
on curve (c). Curve (d), starting at the $TCP$, and curve (b) below the $TCP$
limit the existence regime of ordered and disordered phases respectively.}
\label{fig1}
\end{figure}
\indent
The particularity of the $TCP$ stems from the fact that the 
replica--diagonal fields $Q^{aa}$ become critical in addition to 
the usual replica--overlap fields $Q^{ab}$ with $a\neq b$. 
On approaching the $TCP$ with $T\rightarrow T_{c3}$
and $\mu$ fixed for example, the saddle--point solution 
$\tilde{q}=Q^{aa}_{SP}$, as obtained from Eq.(\ref{one}), develops 
nonanalytical behaviour given by
\begin{equation}
\delta\tilde{q}\equiv\tilde{q}(T)-\tilde{q}(T_{c3})=
A_{\pm}(\pm\tau)^{\frac{1}{2}}+O(\tau)
\label{two}
\end{equation}
with $A_{+}=\sqrt{3}A_{-}\simeq 0.1999$.
This behaviour is at the origin of the crossover from the critical exponents 
$-\alpha=\beta_{q}=\gamma=1$, valid 
for $T_c>T_{c3}$, to the tricritical ones 
$\alpha_3=\beta_{q3}=\gamma_3=\frac{1}{2}$ obtained here for $T\rightarrow 
T_{c3}$. The free energy given above can be cast into the scaling form 
$f=\tau^{2-\alpha}{\it G}(\frac{g}{\tau^{\phi}})$ for the disordered 
phase with $\alpha=-1$ and crossover exponent $\phi=2$.
In the tricritical regime the singular part behaves like $f_{sing}^{TCP}\sim
g^{\frac{3}{2}}$ whence $\alpha_3=\frac{1}{2}$ from above.  
In contrast to crystal--field split spin glasses
\cite{MS} the quartic coefficient $y_4$ of our free
energy,
Eq.(\ref{one}), is nonzero and one obtains the Parisi solution $q(x)=\frac{9}{2y_4}x$ for 
$0\leq x\leq x_1$ and $q(x)=q(1)$ for $x_1\leq x\leq 1$. The plateau 
height is found to satisfy $q(1)=\delta\tilde{q}+O(\delta\tilde{q}^2)$. 
Consequently, plateau and breakpoint scale like $\sqrt{|\tau|}+O(\tau)$ 
at the $TCP$, while linear $\tau$-dependence is reserved 
to $T_c>T_{c3}$. Adapting the notation of 
\cite{DFHS} we express our result for the irreversible response 
$q(1)-\int^{1}_{0}q(x)\sim |\tau|^{\beta_{\Delta}}$ in terms of the 
exponent $\beta_{\Delta 3}=1$ for $T\rightarrow T_{c3}$ and 
$\beta_{\Delta}=2$ for $T\rightarrow T_c>T_{c3}$.\\
\indent
For the Almeida--Thouless line at tricriticality we find
$\frac{H^2}{J^2}=\frac{80}{81}(\frac{2}{3}(1-\frac{\mu_{c3}}{J}
tanh(\frac{3\mu_{c3}}{J})))^{3/2}\tau_{AT}^{3/2}+O(\tau_{AT}^2)$
with $\tau_{{\small AT}}\equiv \frac{T_{c3}-T_{AT}(H)}{T_{c3}}$. Hence we 
obtain the critical exponent $\theta_{3}=\frac{4}{3}$ near $T_{c3}$, while 
$\theta=\frac{2}{3}$ for all $T_c>T_{c3}$. 
These values do not 
satisfy the scaling relation $\theta_{3}=
\frac{2}{\beta_{\Delta_3}}$ with 
$\beta_{\Delta_3}=1+(\gamma_3-\alpha_3)/2$. Along the lines described by D.Fisher 
and H.Sompolinsky \cite{DFHS}, this problem of mean--field exponents will be 
resolved 
below by the renormalization group analysis of the coupling $y_4$ of the 
{\it finite--range and finite--dimensional $ISG_f$}.\\
\indent
The fermionic nature of the $ISG_f$ and of related models on Fock space 
calls for a representation in terms of electron Green's functions. We 
shall see that
even if charge--fields do not occur or do not become critical together 
with the spin--fields, nonanalytic behaviour at the spin--glass 
transition can become not only observable but also intimately 
related to quantities defined in the charge--sector.
We mention our result for the density of states ($DOS$), which is given in 
the disordered phase by
\begin{equation}
<\rho_{\sigma}(\epsilon)>\\
=\frac{ch(\beta\mu)+ch(\beta (\epsilon 
+\mu))}{\sqrt{2\pi\tilde{q}}J[ch(\beta\mu)+exp(\frac{\beta^2 
J^2\tilde{q}}{2})]}
e^{-\frac{(\epsilon +\mu)^2}{2J^2\tilde{q}}}.
\label{three}
\end{equation}
For $T<T_c$ we note that $<\rho>=<\rho>|_{q=0}+O(q^2)$
and the fact that the 
self--consistency equations can be obtained from Eq.(\ref{three}).
The internal energy
$U= \sum \int d\epsilon \epsilon f(\epsilon)<\rho_{\sigma}(\epsilon)>$ 
is found as
$U=-1.996J-1.045\delta\tilde{q}+O(\delta\tilde{q}^2,q^2)$.
Inferring Eq.(\ref{two}) the specific heat is seen to diverge like $c\sim 
|\tau|^{-\frac{1}{2}}$,
confirming thus $\alpha_3=\frac{1}{2}$ on both sides of the $TCP$. 
Transitions at $T_c>T_{c3}$ obey
$\delta\tilde{q}\sim q(1)\sim \tau$ instead. 
The average filling factor $<\nu>\equiv\sum_{\sigma}<n_{\sigma}>$ obeys the 
relation 
$<\nu>=1+tanh(\beta\mu)(1-\tilde{q})$ 
(we find that this relation is 
invariant under replica symmetry breaking). 
For fixed $\mu$ this implies
that $<\nu>$ shows nonanalytical $\sqrt{|\tau|}$--behaviour near the $TCP$,
and the same holds for the electronic density of states. Collecting the 
results by
\begin{equation}
\frac{d<\rho_{\sigma}(\epsilon)>}{dT}\sim 
\frac{d<\nu>}{dT}
\sim\frac{d\tilde{q}}{dT}
\sim |\tau|^{-1/2}
\end{equation}
we conclude that all these quantities diverge with the 
MF--exponent $\alpha_3$
of the specific heat both from above and below $T_{c3}$.
The local susceptibility 
$\chi=\frac{1}{J(1+\tau)}(1+\delta\tilde{q}-\int^{1}_{0}q(x))$ has a 
{\it divergent slope for $T\downarrow T_{c3}$} but a finite one for $T\uparrow 
T_{c3}$ due to our result $q(1)=\delta\tilde{q}+O(\delta\tilde{q}^2)$.    
Inferring 
Eq.(\ref{two}) one finds $d\chi/dT\sim
\tau^{-\frac{1}{2}}\theta(\tau)+O(\tau)$.
It is thus the nonanalytic change of the fermion concentration, 
which yields in combination with the ubiquituous
replica--overlap critical fluctuations the special tricritical behaviour 
as the $TCP$ is approached. What happens if one wishes to 
consider $\mu$ as a function of given average filling $<\nu>$?
Again the exact relation given above, which holds true if $\mu(<\nu>)$ is
constrained to be nonrandom, 
shows that now the chemical potential will acquire 
$\sqrt{|\tau|}$--corrections. The other conclusions remain unchanged.\\
\indent
The fermionic picture also allows to study correlations of 
the statistical fluctuations of the density of states.
The $DOS$--cumulant 
$<\delta\rho_{\sigma}^a(\epsilon)\delta\rho_{\sigma}^{b\neq 
a}(\epsilon^{\prime})>$ with $\delta\rho\equiv\rho\hspace{1mm} -<\rho>$
is sensitive to the spin glass transition.
We obtain  
$<\delta\rho^a(0)\delta\rho^b(0)>=
\frac{A(0,0;\mu)}{J^2}q^2+O(q^3)$ with 
$A(0,0;0\leq|\mu|\leq|\mu_{c3}|)$ decreasing monotonuously 
from the maximum value $0.6592$ at half filling to $0.3613$ at the 
tricritical point. 
Hence this cumulant, being defined in the charge sector, 
decays to zero at the spin--glass transition quadratically in $\tau$ for 
$T_c>T_{c3}$ and linearly at the $TCP$. 
Beyond this point it is discontinuous at the transition. 
The {\it DOS}--cumulant also feels replica--symmetry breaking and becomes 
proportional to $q^2(x)$. In a nonvanishing magnetic field the 
$DOS$--cumulant becomes nonzero also above $T=T_c(H=0)$.\\
\indent
As line (b) of Fig.1 merges with line (a) at the $TCP$, the replica--diagonal
fields $Q^{aa}$ become critical together with the off--diagonal fields,
a phenomenon very unusual for 
classical thermal spin glass transitions 
and so far only known to occur in special limits \cite{Moore}. In 
addition $Q^{aa}$ appears linearly in the Lagrangian and hence plays a 
special role. These crucial features are surprisingly shared by the thermal
tricritical theory and the $T=0$--quantum theory for metallic spin glass--
\cite{SRO} 
and for the transverse field Ising spin glass transitions \cite{RSY}.
These crosslinks with $T=0$--quantum phase transitions with irrelevant
quantum dynamics (for dynamic exponents $z>2$) are best appreciated in 
field theory. We derived the Lagrangian for the tricritical and {\it finite
range $ISG_f$} and a Lagrangian of the same structure
is obtained for generalized models (eg with a transport
mechanism) at {\it finite} temperature by integrating out the
dynamical degrees of freedom
\begin{eqnarray}
L&=& \frac{1}{t}\int d^dx{\large[}\frac{r\kappa_1}{(\kappa_2)^2}\sum Q^{aa} 
+ \frac{1}{2}\sum Q^{aa}(-\nabla^2+u)Q^{aa}
\nonumber\\
&+&\frac{1}{2}Tr^{\prime}(\nabla Q^{ab})^2 
- \frac{1}{t}\sum^{\hspace{.6cm}\prime} Q^{aa}Q^{bb}  
-\frac{\kappa_1}{3}\sum (Q^{aa})^3\nonumber\\
&-&\frac{\kappa_3}{3}
Tr^{\prime}Q^3
-\kappa_2\sum^{\hspace{.5cm}\prime} 
Q^{aa}Q^{ab}Q^{ba}+\frac{y_4}{4}\sum^{\hspace{.5cm}\prime} (Q^{ab})^4 
{\large],} 
\label{Lagr}
\end{eqnarray} 
where 
$4(\frac{\kappa_1}{t})^{(0)}=(\frac{\kappa_2}{t})^{(0)}=
(\frac{\kappa_3}{t})^{(0)}
=\frac{3^3}{2}$ 
and $u^{(0)}=0$ denote the bare coefficients at tricriticality.  
One fourth order term relevant for replica symmetry breaking is kept.
Replicas under 
$\sum^{\prime}$ or $Tr^{\prime}$ are distinct. The $Q^{aa}Q^{bb}$--coupling 
is renormalization group generated as in the metallic quantum spin 
glass and leads to a common upper critical dimension $d_c^{(u)}=8$.
A shift of the fields removing the redundant $(Q^{ab})^2$-mass term was 
performed at each step of our one--loop renormalization group analysis.
Since $Q^{aa}$ and $Q^{ab}$--fields may possess different anomalous 
dimensions, they are kept separately in the Lagrangian. Apart from the 
striking similarities between the tricritical $ISG_f$ Lagrangian and the 
metallic quantum case there are also some important differences:\\
\indent
i) Time--integrals are absent, each $Q$--field may be viewed as
$Q_{\omega=0,\omega=0}$(x).\\
\indent
ii)Unlike the metallic model 
the linear term is time--independent.\\
\indent
iii)
Three relevant cubic couplings appear instead of one in the metallic case.\\
\indent
iv) The dangerously irrelevant (DI) $u\int d\tau (Q^{aa}_{\tau\tau})^2$ quantum
mechanical interaction of the metallic case \cite{SRO} turns into a relevant 
mass term for the $ISG_f$.\\
\indent
While in case of the metallic $T=0$--transition the $u$--interaction was seen
to render the nonlinear susceptibility $\chi_{nl}$ less divergent than 
the spin glass susceptibility $\chi_{SG}$, the $ISG_f$ shows 
$\chi_{nl}\sim\chi_{SG}$ due to property $iv)$.
This was confirmed by our MF calculation
of the small field behaviour of the magnetization m which yields to $O(H^3)$
$m = \frac{H}{J} - 
(\frac{27}{2}(1-\frac{\mu_{c3}}{J}th(\frac{3\mu_{c3}}{J})))^{\frac{1}{2}} 
\frac{H^3}{J^3\sqrt{\tau}}$ 
for $T>T_{c3}(H=0), \mu=\mu_{c3}$ near the tricritical point.
DUE to $\chi^{-1}_{SG}\sim\tau$ near 
$T_c>T_{c3}$ and $\chi^{-1}_{SG}\sim\sqrt{|\tau|}$ 
the $Q$--propagator $1/(k^2+m^2)$ 
yields the MF 
correlation length exponent $\nu_3=\frac{1}{4}$ at the $TCP$ and
$\nu=\frac{1}{2}$ in the continuous regime above. 
As for the metallic
$T=0$--transition the MF--exponents violate hyperscaling even in the upper
critical dimensions $d_{c3}^{(u)}=8$ ($TCP$ of $ISG_f$). 
The $\frac{1}{t^2}Q^{aa}Q^{bb}$--term is at the origin of this violation for 
both the 
tricritical point of the $ISG_f$ and for the $T=0$--metallic spin--glass 
transition. In the metallic case ($z=4$) the replacement of dimension d by the 
'quantum'--mechanical dimension $d_{qm}=d+2z-\theta_t$ 
yields modified hyperscaling relations which are satisfied by
MF--exponents in $d=d_c^{(u)}=8$.
The same result is obtained for the tricritical $ISG_f$ by replacing
$d\rightarrow d-\theta_t$. In both cases $\theta_t$ denotes the dimension
of the DIC t.\\         
\indent
We studied tricritical fluctuation 
in a 1-loop
renormalization group analysis for the Lagrangian 
(\ref{Lagr}). Anomalous dimensions $\eta$ and $\tilde{\eta}$ are 
introduced to account for $Q^{a\neq b}$-- and for $Q^{aa}$--fluctuations,
respectively. These exponents and the one for the $DIC$ t 
are given by
$\eta=\frac{2\kappa^2_2}{(1+u)^2},
\tilde{\eta}=\frac{2\kappa^2_1}{(1+u)^3},
\theta_t=2.$
We obtain the following RG flow equations 
\begin{eqnarray}
\frac{dr}{dl}&=&(\frac{d}{2}-11\kappa^2_1+16\kappa_1\kappa_2+6\kappa^2_2)r-
\kappa^2_2,\nonumber\\
\frac{d\kappa_1}{dl}&=&\frac{\epsilon}{2}\kappa_1+9\kappa_1^3,
\frac{d\kappa_2}{dl}=(\frac{\epsilon}{2}+6\kappa^2_2-
\kappa_1^2+16\kappa_1\kappa_2)\kappa_2\nonumber\\
\frac{d\kappa_3}{dl}&=&(\frac{\epsilon}{2}+9\kappa^2_2)\kappa_3,
\frac{du}{dl}=2(1-\kappa_1^2)u-4\kappa_1^2+4\kappa_1\kappa_2,
\end{eqnarray}
where $\epsilon=8-d$. Above $d=8$ the Gaussian fixed point is 
stable, while for $d<8$ a runaway flow to strong coupling occurs, reminding 
of the one observed for the quantum phase transition of 
metallic models \cite{SRO} and of transverse field models \cite{RSY}.
The runaway flow is expected within the first order regime, $u^{(0)}<0$,
but a strong coupling tricritical fixed point limiting the known second 
order regime is still to be found.\\ 
\indent
The RG for the $DIC$ $y_4$ showed that its long--distance behaviour 
is dominated by a $\kappa^4$--contribution (like in \cite{DFHS} but) for 
$d_c^{(u)}=8<d<10$.
This leads to the modified MF exponent 
$\theta_3=\frac{8}{d-4}$, which satisfies the scaling relation 
$\theta_3=2/\beta_{\Delta_3}$ in $d_{c3}^{(u)}=8$ 
and reduces to the MF--result 
in $10$ dimensions.
The dimensional shift by $2$ in comparison with Ref.\cite{DFHS} is due to 
coupling $t$.\\
\indent
For given {\it nonrandom} chemical potential the frustrated spin interaction 
generates weakly nongaussian statistical fluctuations of the fermion filling
and vice versa, half filling exempted. Imposing instead a Gaussian 
$\delta\mu$--distribution one finds the present problem mapped onto
onto the $Q$--static approximation \cite{ROMB} of a metallic Ising spin 
glass in the limit of electron hopping range zero. 
This limit turns random hopping 
matrix elements into random site--local energies, 
which are equivalent to the
fluctuating chemical potential and render the metallic Ising spin glass 
classical and static. The main new effect of $\mu$--randomness is the 
generation of a {\it classical} $T=0$--transition at 
$<(\delta\mu)^2>_c=(\frac{16J}{3\pi})^2$ for
$<\mu>=0$.\\ 
\indent
We find that the random--$\mu$ $ISG_f$--model and the metallic Ising
spin glass show tricritical behaviour on the $T=0$--axis as well as in their
thermal transitions with discontinuous low-- and high--filling--regimes.
At $T=0$ we find the {\it tricritical point of the metallic
spin glass} with gaussian random hopping at
\begin{equation}
E_F = \sqrt{1-\sqrt{\frac{5}{8}}}E_0 \quad,\quad  J_c=\frac{3\pi
E_0}{32}[1-\frac{E_F^2}{E_0^2}]^{^{-\frac{3}{2}}},
\end{equation}
where $2E_0$ denotes the width of the semielliptic
electronic band. 
Quantum dynamical corrections 
can be approximated by a generalized
Miller--Huse method \cite{MilHu}.
Details of the present work will be given elsewhere.\\
\indent
We thank J. T. Chalker, Th. Garel, and Subir Sachdev for helpful comments. Our research
was supported by grant $Op28/4$--$1$ and by the $SFB410$ of 
the DFG.\\

\end{multicols}

\begin{references}
\bibitem{Sachdev1} S. Sachdev and J. Ye, Phys. Rev. Lett. {\bf 70}, 3339 
(1993). 
\bibitem{Sachdev2} J. Ye, S. Sachdev, and N. Read, Phys. Rev. Lett. {\bf 70},
4011 (1993).
\bibitem{Georges} A. Sengupta and A. Georges, Phys. Rev. B {\bf 
52} 10295 (1995).             
\bibitem{RSY} N. Read, S. Sachdev, J. Ye, Phys. Rev. B{\bf 52}
384 (1995).
\bibitem{SRO} S. Sachdev, N. Read, and R. Oppermann, Phys. Rev. B {\bf 52}
10286 (1995).
\bibitem{Maple} M. B. Maple {\it et al.}, J. Low Temp. Phys. {\bf 95}, 
225 (1994).
\bibitem{Loehn} H. von Loehneysen {\it et al.}, Phys. Rev. Lett. {\bf 
72}, 3262 (1994).
\bibitem{LaSrCuO} F. C. Chou, N. R. Belk, M. A. Kastner, R. J. Birgeneau,
and A. Aharony, Phys. Rev. Lett. {\bf 75}, 2204 (1995). 
\bibitem{BEG} M. Blume, V. J. Emery, and R. B.
Griffiths, Phys. Rev. A{\bf 4}, 1071 (1971).
\bibitem{MS} P. J. Mottishaw, D. Sherrington, J. Phys.
C {\bf 18}, 5201 (1985).
\bibitem{DFHS} D. S. Fisher and H. Sompolinsky, Phys.Rev.Lett.{\bf 54}, 
1063 (1985).
\bibitem{Moore} J. E. Green, A. J. Bray, M. A. Moore, J. Phys. A {\bf 15},
2307 (1982).
\bibitem{ROMB} R. Oppermann and M. Binderberger, Ann. Phys. {\bf3}, 494 
(1994).
\bibitem{MilHu} D. Huse and J. Miller, Phys. Rev. Lett. {\bf 70}, 3147 
(1993). 
\bibitem{LawSa} I. D. Lawrie and S. Sarbach, in {\it Phase Transitions 
and Critical Phenomena} edited by C. Domb and J. L. Lebowitz (Academic Press,
London, 1984), p.1

\end{references}
\end{document}